\documentclass[twocolumn,aps,pra,nofootinbib,floatfix,longbibliography,groupedaddress]{revtex4-2}
\usepackage{amsmath,bbm}
\usepackage{graphicx}
\usepackage{multirow}
\usepackage{hyperref}
\usepackage{amsfonts}
\usepackage{amsbsy}
\usepackage{xcolor}
\usepackage{soul}
\usepackage{mathtools}
\usepackage{bm}

\usepackage{bbold}

\renewcommand{\vec}[1]{\mathbf{#1}}

\newcommand{\Pc}{\mathcal{P}}
\newcommand{\cbE}{\boldsymbol{\mathbf{\cal E}}}

\newcommand{\ket}[1]{\ensuremath{\left\vert #1 \right\rangle}}
\newcommand{\unitvec}[1]{\hat{\mathbf{{#1}}}}

\newcommand{\pol}{\hat{\bf e}}
\newcommand{\rv}{{\bf r}}

\newcommand{\beq}{\begin{equation}}
\newcommand{\eeq}{\end{equation}}
\newcommand{\bea}{\begin{eqnarray}}
\newcommand{\eea}{\end{eqnarray}}
\newcommand{\BEQAL}{\begin{align}}
\newcommand{\EEQAL}{\end{align}}
\newcommand{\EQREF}[1]{Eq.~(\ref{#1})}

\newcommand{\comment}[1]{{}}

\newcommand{\commentout}[1]{{}}

\begin{document}
\title{Wavefront engineering with atomic Huygens' surfaces made simple}

\date{\today}
\author{K.~E.~Ballantine}
\author{J.~Ruostekoski}

\affiliation{School of Physics and Astronomy, Lancaster University, Lancaster, LA1 4YB, United Kingdom}

\begin{abstract}
Atomic arrays provide a cooperative platform for light-mediated interactions in nonlocal metasurfaces which can operate in the quantum regime without nanofabricated optical elements or intrinsic material absorption. 
We present a minimal atomic implementation to harness wavefront engineering and create a near-ideal  intrinsic atomic Huygens' surface. A simple bilayer square lattice with a two-atom unit cell, not limited to small lattice spacings,
provides near-unity reflectionless transmission and a full $2\pi$ phase control of transmitted light without the localized site-dependent level shifts of previous proposals. This simplification is achieved through nonlocal radiative excitations that provide the familiar electric-dipole response with even radiation parity as well as an odd-parity radiation channel whose leading multipolar interpretation combines magnetic-dipole and electric-quadrupole contributions in atoms exhibiting only electric dipole transitions. This eliminates the need to simultaneously excite pure electric-dipole and magnetic-dipole responses with exactly matched radiation. We derive a two-mode model for this response and show that it accurately captures full coupled-dipole simulations near the optimal lattice spacings. Because of the wide applicability of coupled-dipole models, the relevance of our microscopic analysis extends beyond atomic physics to nanostructured metasurfaces. Moreover, the same platform also connects directly to established subradiance-protected single-photon storage protocols, suggesting a route toward combining wavefront engineering, storage, and retrieval in a simple atomic bilayer.

\end{abstract}

\maketitle

\section{Introduction}

Optical metasurfaces are thin, effectively two-dimensional metamaterial layers that can control the phase, amplitude, and polarization of light~\cite{Yu14,Chen2016_review,Luo18,Qiu21_review}. An important example of a metasurface is the Huygens’ surface that can provide full phase control in transmission while suppressing reflection~\cite{Pfeiffer13,Decker15,Yu15,Chong15,Shalaev15}.

Huygens' principle states that each point in a diffracting beam acts as a secondary source, all of which interfere to construct the observed propagation~\cite{Huygens}. A more rigorous formalism, due to Love, describes these sources as crossed electric and magnetic dipoles~\cite{Love1901}. This electric-magnetic description explains the asymmetry between the forward and backward directions: the two radiation channels can interfere constructively in transmission and destructively in reflection.  Since any beam can be described by these fictitious sources, in turn a physical implementation of a subwavelength array of crossed electric and magnetic dipoles with controllable phase can, in principle, generate arbitrary wavefronts~\cite{Pfeiffer13,Decker15,Yu15,Chong15,Shalaev15}.

Ordered subwavelength two-dimensional planar arrays of atoms offer a microscopic and highly tunable route to harnessing cooperative light-mediated interactions and manipulating light~\cite{Jenkins2012a}. This provides a versatile platform for nonlocal metasurface physics~\cite{Ruostekoski2023} with quantum-optical control that is challenging to achieve in conventional nanostructured metasurfaces~\cite{Solntsev21}.
Atom trapping technology in periodic arrays has advanced rapidly~\cite{Rui2020,Srakaew22,bluvstein2022,Ferioli21,Lester15,Barredo18,Mello19,Kim16,Cooper18}, and optical lattices with unit filling can be experimentally realized with hundreds of thousands of atoms~\cite{Shao2024}.
Transmission experiments have revealed giant subradiant collective radiative excitations that extend over the Mott-insulator state in an optical lattice~\cite{Rui2020} and demonstrated coherently manipulated optical switching~\cite{Srakaew22}. Theoretically, 
the range of prospective applications of planar atomic arrays is unusually broad, including  single-photon storage~\cite{Ballantine21quantum,Rubies22,Ballantine22bilayer},  coupling to long-wavelength Rydberg excitations~\cite{Bekenstein2020,Moreno2021,Zhang2022}, transduction~\cite{Milczewski26}, interplay with recoil effects~\cite{Eltohfa25}, and manipulating interactions with spin statistics~\cite{Orioli19}---to name just a few examples. There is also an increasing number of studies in bilayer arrays~\cite{Grankin18,Yoo2018,Guimond2019,Alaee20,Yoo2020,Ballantine20Huygens,Ballantine21wavefront,Ballantine21quantum,Ballantine22str,Ballantine22bilayer,Pedersen23,Nefedkin23}.  

Previous studies have shown that a Huygens' surface can be formed from cooperatively responding atoms in non-Bravais lattices using localized site-dependent control of the atomic resonance shifts within the unit cell~\cite{Ballantine20Huygens,Ballantine21wavefront,Ballantine21quantum,Ballantine22str}. 
A Huygens' surface operation is possible even with each atom exhibiting only electric dipole transitions, since nonlocal collective radiative excitations in the presence of more complex unit cells of multiple atoms can exhibit effective magnetic dipole moments.
Configurations based on local level shift control in non-Bravais lattices are experimentally achievable, e.g.\ by utilizing ac Stark shifts of lasers for realistic atomic transitions~\cite{Ballantine22str}, but they require accurate spatial control and laser alignment at the level of the unit cell.
Other proposed schemes of wavefront control in atomic and quantum emitter arrays include atomic polarizers~\cite{Wang17,Bassler23,Wang26} and a
metalens with layers of emitters in concentric rings of variable phase shifts and emitter spacings, e.g., in a solid-state substrate~\cite{Andreoli25}. 

\begin{figure}[t]
 \centering
  \includegraphics[width=\columnwidth]{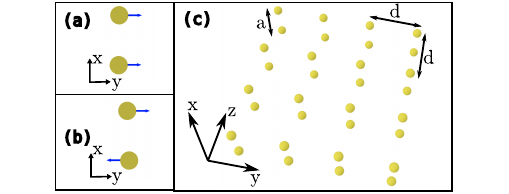}
 \caption{A minimal atomic Huygens' unit cell construction with (a) symmetric mode polarization and (b) antisymmetric mode polarization on a single two-atom unit cell. (c) Geometry of a bilayer lattice with separation $d$ in the $yz$ plane and separation $a$ in the $x$ direction. Collective modes consist of uniform polarization across the $yz$ plane, with the corresponding symmetry between atoms at $x=0$ and $x=a$. The lattice can be considered as two regular square arrays, which we label by $A$, $B$. separated in the $x$ direction, or alternatively as a single square array with two atoms per unit cell.  }
 \label{fig:fig1}
\end{figure}

Here we propose a minimal atomic Huygens' surface with a simple bilayer construction of two square arrays of two-level atoms without externally engineered level shifts or complex in-plane unit-cell structures. In particular we consider an array with each unit-cell occupying a square lattice with spacing $d$ in the $yz$ plane, and consisting of two atoms with separation $a$ in the $x$ direction (see Fig.~\ref{fig:fig1}). The key simplification is that a Huygens' surface requires not distinct electric- and magnetic-dipole resonances, but two radiation channels with different forward/backward parity. In a bilayer atomic array, these channels are supplied intrinsically by the symmetric and antisymmetric collective excitations of a two-atom unit cell, as the latter has the radiation symmetry of a magnetic-dipole/electric-quadrupole superposition. Their backward-scattered fields can therefore cancel, while forward-scattered fields combine with the incident field to produce high-transmission phase control.

For normal incidence, the propagating laser is phase matched to two collective excitation eigenmodes that have uniform phase in the transverse plane and are symmetric or antisymmetric, respectively, between the two layers. We derive a two-mode model for these collective excitations and show that it captures the response of the full many-atom dynamics near the optimal lattice spacings. At interlayer spacings $a=\lambda/4$ and $5\lambda/4$, the incident field drives the two modes with equal strength, and for the corresponding in-plane spacings $d\simeq0.91\lambda$ and $0.96\lambda$, respectively, their collective resonances overlap. The resulting bilayer acts as a Huygens' surface with near-unity transmission and with a transmission phase that winds continuously by $2\pi$ and is controlled by the laser detuning alone. We then demonstrate its operation as a tunable wave plate: if one linear polarization component couples to the array, varying the detuning changes its transmitted phase relative to the orthogonal component, providing continuous control of the output Stokes vector.

Coupled point-dipole models apply to a wide range of metasurface systems, where they have successfully described experimentally observed nonlocal cooperative phenomena~\cite{Jenkins17,Jenkins18,Papasimakis19}. The relevance of our results therefore extends beyond atomic physics, providing microscopic insight into the operation of Huygens' metasurfaces more generally.

The simple two-level scheme can be extended to the isotropic  $J=0\rightarrow J'=1$ transition which is relevant for alkaline-earth-metal and rare-earth-metal atoms with flexibility in experimental control~\cite{Daley08,Fukuhara09,Ye08,Olmos13,Ballantine22str}  and applications in atomic clocks~\cite{Bothwell19}.
For this isotropic transition the atomic layers also support strongly subradiant collective modes with dipoles pointing out of plane~\cite{Facchinetti16,Facchinetti18,Ballantine20ant}. Such modes can be coupled to the Huygens' surface modes using ac-Stark shifts, connecting the Huygens' surface scheme to the established subradiant single-photon storage protocol of Ref.~\cite{Ballantine22bilayer}, 
suggesting a common platform for wavefront shaping, storage, and retrieval.

\section{Basic Theory}

We consider both two-level atoms and the more general $J=0\rightarrow J'=1$ atomic level structure. In atom $j$, the ground level $|g_{j}\rangle$ is coupled to the excited level $|e_{j\nu}\rangle$. For two-level atoms,  the excited level index $\nu$ is fixed, while for the $J=0\rightarrow J'=1$ transition the index $\nu$ runs over the sublevels.   The main discussion is focused on the two-level case and the isotropic  $J=0\rightarrow J'=1$ atomic transition is analyzed in Appendix~\ref{sec:iso}.

The atoms interact with a weak coherent incident light field $\boldsymbol{\mathcal{E}}$, with frequency $\omega=c |{\bf k}|=ck=2\pi c/\lambda$ and wavevector ${\bf k}$, and also with light scattered from other atoms in the array. 
The positive frequency component of the field amplitude is $\boldsymbol{\mathcal{E}}{}^{+}(\textbf{r})=\mathcal{E}(\vec{r})\exp{(ikx)}\unitvec{e}_{\rm in}$. We examine the case of an incident plane wave, $\mathcal{E}(\vec{r})=\mathcal{E}_0$, a constant, and Gaussian beam, with beam waist $w$ comparable to the size of the lattice. Without loss of generality, we choose $\mathcal{E}_0$ to be real. 
 
The positive frequency component of the atomic polarization density is denoted by $\hat{\textbf{P}}{}^+(\textbf{r})=\sum_j \hat{\bf d}^{(j)}\delta(\vec{r}-\vec{r}_j)$,  where $\hat{\bf d}^{(j)}$ and $\vec{r}_j$ are the dipole operator and position of atom $j$, respectively. 
In the following, we refer to slowly varying amplitudes  for both atom and light field quantities, where the rapid oscillations at the laser frequency have been factored out, such as $\exp(-i \omega t)$ from $\hat{\textbf{P}}{}^+(\textbf{r})$.
For a two-level atom, we write the dipole moment $\hat{\bf d}^{(j)}\rightarrow \mathcal{D}\mathcal{P}^{(j)} \pol_d={\bf d} \mathcal{P}^{(j)}$, where  $\mathcal{P}^{(j)}$ is the polarization amplitude on atom $j$. 

In the low light intensity limit, which neglects terms beyond first order in the atomic excitation and  driving field amplitudes, the dynamics of coherently driven stationary atoms with a single electronic ground level can be exactly described by classical coupled-dipole equations~\cite{Javanainen1999a,Lee16}.
This regime has been shown  to accurately describe dense gas optical responses in experiments~\cite{Rui2020,Vatre2026}.
 In the two-level case, we then obtain the low light intensity dynamics in the rotating wave approximation~\cite{Lee16}, 
\begin{equation}
\label{eq:eom}
\dot{\vec{b}}=i(\mathcal{H}+\delta \mathcal{H})\vec{b}+ \vec{f},
\end{equation} 
where $b_j=\Pc^{(j)}$, $f_j=i {\cal R} (\vec{r}_j)$  for $j=1,\ldots,N$, and Rabi frequency $ {\cal R}=\mathcal{D}\pol_d^*\cdot\boldsymbol{\mathcal{E}}/\hbar$ (note in the following we drop the $(+)$ notation for simplicity, and assume all fields refer to the positive frequency component). The diagonal elements of $\mathcal{H}$ and  $\delta \mathcal{H}$ are  $i\gamma$ and $\Delta^{(j)}$, respectively. Here $\gamma$ is the single atom resonance linewidth, and $\Delta^{(j)}=\Delta+\delta^{(j)}$ is the detuning of the laser frequency from the resonance of atom $j$, consisting of the detuning $\Delta=\omega-\omega_0$ from the unperturbed single-atom resonance $\omega_0$ as well as any applied level shifts $\delta^{(j)}$, e.g., from the ac Stark shift of additional laser or microwave fields~\cite{gerbier_pra_2006}. The off-diagonal elements of $\delta \mathcal{H}$ vanish and those of $\mathcal{H}$ are (for $\xi=6\pi\gamma/k^3$)
\begin{equation}
i\xi \unitvec{e}_d^{\ast}\cdot \mathsf{G}(\vec{r}_j-\vec{r}_k)\unitvec{e}_d,
\end{equation}
for $j\neq k$, with $\mathsf{G}$ the standard dipole radiation kernel~\cite{Jackson}
\beq
{\sf G}_{\nu\mu}({\bf r}) =
\left[ {\partial\over\partial r_\nu}{\partial\over\partial r_\mu} -
\delta_{\nu\mu} \boldsymbol{\nabla}^2\right] {e^{ikr}\over4\pi r}
-\delta_{\nu\mu}\delta({\bf r})\,.
\label{eq:GDF}
\eeq

The response of the lattice can be frequently understood by examining the eigenmodes of the symmetric but non-Hermitian matrix $\mathcal{H}$, given by $\mathcal{H}\vec{v}_n=(\delta_n+i\upsilon_n)\vec{v}_n$~\cite{Rusek96,JenkinsLongPRB,Jenkins2012a,Castin13,Jenkins_long16,Ruostekoski2023}, where the eigenvalues $\delta_n+i\upsilon_n$ of the $n^{\mathrm{th}}$ mode consist of the collective line shift $\delta_n$ and the collective linewidth $\upsilon_n$. The eigenmodes are generally not orthogonal, but satisfy biorthogonality relations $\vec{v}_j^T \vec{v}_k=\delta_{jk}$, except for some possible cases of $\vec{v}_j^T \vec{v}_j=0$~\cite{Ruostekoski2023}, which do not appear in our numerics. We can then define~\cite{Facchinetti16}
\begin{equation}
    L_j=\frac{|\vec{v}_j^T\vec{b}|^2}{\sum_k|\vec{v}_k^T\vec{b}|^2},
\end{equation}
as a measure of the occupation of the eigenmode $\vec{v}_j$ for a given state $\vec{b}$.

\section{Bilaryer Array}

\subsection{Geometry}

We now turn to the specific geometry of a bilayer array, consisting of two layers, which form a regular square array with spacing $d$ in the $yz$ plane, separated by interlayer separation $a$ in the $x$ direction, as illustrated in Fig.~\ref{fig:fig1}. This lattice can equivalently be thought of as a regular square lattice in the $yz$ plane with two atoms per unit-cell.
Here, we consider a lattice of $n\times n$ atoms in the $yz$ plane with two layers in the $x$ direction, for a total of $N=2n^2$ atoms. 
In the main body we consider a two-level transition with the dipoles oriented along the $y$ direction, in the lattice plane, $\unitvec{e}_d=\unitvec{e}_y$, while in Appendix~\ref{sec:iso} we consider an isotropic $J=0\rightarrow J'=1$ transition. The light is incident normal to the array, in the $x$ direction, with wave vector $\mathbf{k}=k\unitvec{x}$ and in-plane polarization $\unitvec{e}_\mathrm{in}=\alpha\unitvec{e}_y+\beta\unitvec{e}_z$. 

Assuming a unidirectional input field in the forward direction, the reflected light consists solely of the scattered field. 
While for a single layer the forward and backward scattered fields are identical, here interference between the two layers breaks this symmetry. 
We write the scattered fields as $\mathbf{E}^{(f,b)}$, where $f$, $b$, refer to the forward and backward directions, respectively.
Then the total backward-propagating field is $\mathbf{E}_{\mathrm{tot}}^{(b)}=\mathbf{E}^{(b)}$. However, in the forward direction the scattered field interferes with the incident field, and the total propagating field transmitted through the lattice is $\mathbf{E}^{(f)}_{\mathrm{tot}}=\mathbf{E}^{(f)}+\cbE(\mathbf{r})$.

To demonstrate the efficiency of the Huygens' surface numerically for a finite lattices, the forward transmitted filed $\vec{E}_{\mathrm{tot}}^{(f)}$ is calculated by adding the incident field to the transmitted field calculated in the forward scattering approximation~\cite{Chomaz12,Javanainen17,Facchinetti18,Javanainen19,Ruostekoski2023}, while the reflected field consists simply of the scattered field in the same approximation:
\begin{align}
\label{eq:Ef}
\epsilon_0 \vec{E}_{\mathrm{tot}}^{(f)} &= \left[\epsilon_0 \mathcal{E}_0\unitvec{e}_\mathrm{in}+\frac{ik}{2\mathcal{A}}\sum_{j=1}^N\mathcal{D}\mathcal{P}^{(j)}e^{-ikx_{j}}\unitvec{e}_d\right]e^{ikx}, \\
\label{eq:Eb}
\epsilon_0 \vec{E}_{\mathrm{tot}}^{(b)} &= \frac{ik}{2\mathcal{A}}\sum_{j=1}^N\mathcal{D}\mathcal{P}^{(j)}e^{ik(x_{j}-x)}\unitvec{e}_d
\end{align}
where $\mathcal{A}$ is the total area of the lattice, and we assume $\unitvec{e}_\mathrm{in}$, $\unitvec{e}_d$ are both in the $xy$ plane. In particular, when $\unitvec{e}_\mathrm{in}=\unitvec{e}_d$ there is only one relevant polarization and we can define the scalar transmitted field $E^{(f)}$ by $\vec{E}^{(f)}=E^{(f)} e^{ikx}\unitvec{e}_d$. Here, the approximation is valid when $x$ clearly exceeds $d$ and $ x \ll \sqrt{\mathcal{A}}$, where $\mathcal{A}=n^2 d^2$. In this range, the exact value of $x$ is irrelevant for transmission up to an overall phase factor.

\subsection{Two-mode model}
\label{sec:tmm}

\begin{figure}[t]
 \centering
  \includegraphics[width=\columnwidth]{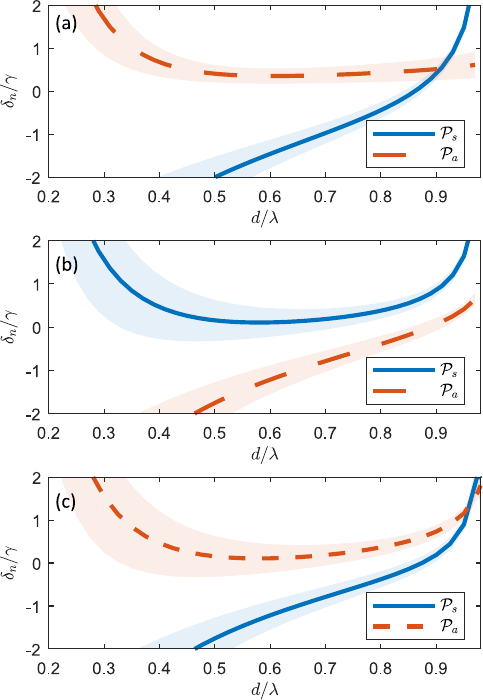}
 \caption{Collective line shifts (solid lines) and linewidths (shaded regions) of phase-uniform symmetric or antisymmetric eigenmodes for varying in-plane spacing $d$ ($n=30)$. We consider cases (a) $a/\lambda=0.25$, (b) 0.75, (c) 1.25, where the two modes are equally driven with equal linewidths. (a) Line shifts are equal for $d\approx0.91\lambda$. (b) Line shifts are not equal for any $d$. (c) Line shifts are equal for $d\simeq 0.96\lambda$. (a) and (c) satisfy the ideal Huygens' condition in the large lattice limit. }
 \label{fig:evals}
\end{figure}

In order to numerically study the scattering of light from the bilayer array, we solve the full set of linear equations, Eq.~(\ref{eq:eom}). However, for $y$ polarized excitation by a plane-wave, or a Gaussian beam with beam-waist comparable to the lattice size, we find the results can be understood by a simple model which considers only two eigenmodes of $\mathcal{H}$ that are phase-matched to the field, along with their eigenvalues~\cite{Facchinetti16,Facchinetti18,Ballantine20Huygens,Ballantine21PT}. These modes consist of a phase-uniform symmetric mode, where all dipoles oscillate with approximately equal phase for both layers, and an anti-symmetric mode, with uniform phase in the $yz$ plane, but a $\pi$ phase shift between the first layer at $x=0$ and the second layer at $x=a$. The behavior of collective linewidths and line shifts of these eigenmodes, obtained by solving the full numerics of the atomic system in Eq.~(\ref{eq:eom}), is shown in Fig.~\ref{fig:evals}. For an ideal infinite lattice, these modes also have equal polarization amplitude on every atom, while for a finite lattice the amplitude falls off towards the edge. The two-mode model is derived in Appendix~\ref{sec:tmmapp}.

\begin{figure}[t]
 \centering
  \includegraphics[width=\columnwidth]{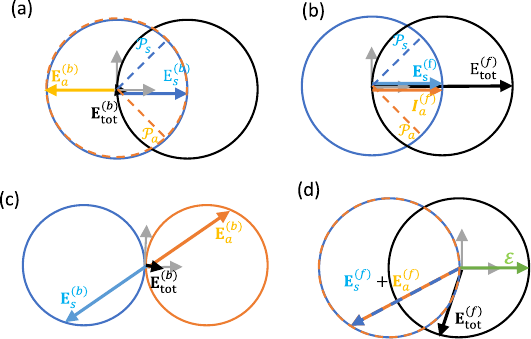}
\caption{Interference of light in forward and backward directions due to differing mode symmetry. As the phase and amplitudes of the fields change they trace out their respective circles in the complex plane. In (a,b) we show the effect of varying the phase of the symmetric and anti-symmetric modes. When these modes have a $\pi/2$ phase difference, they interfere destructively in the backward direction, and constructively in the forward direction. If the phase varies due to the layer spacing, then the phase of the field $\mathbf{E}_{s}^{(b,f)}$ also changes, leading to different interference. (c,d) At $a=\lambda/4$ or $5\lambda/4$ the phase difference is constant, and the response is controlled as the detuning is varied through the resonance. (c) While the amplitude and phase of the scattered field from each mode vary, they interfere destructively with $\mathbf{E}_s^{(b)}=-\mathbf{E}_a^{(b)}$, leading to zero reflection. (d) In the forward direction we have $\mathbf{E}_s^{(f)}=\mathbf{E}_a^{(f)}$. However, interference with the incident field $\mathcal{E}$ means the total field $\mathbf{E}_{\mathrm{tot}}^{(f)}$ has constant amplitude and full $2\pi$ phase variation.}
 \label{fig:sym}
\end{figure}

The differing symmetry of these two modes allows for the full transmission through the atomic metasurface with zero reflection and full control of the phase of the transmitted field. 
This demonstrates the central point: the antisymmetric mode need not be a pure magnetic dipole, as it only needs the same forward/backward scattering parity as the magnetic channel in a conventional Huygens' surface construction~\cite{Love1901}.
We can further decompose the scattered field into contributions of the two modes, $\vec{E}^{(f,b)}=\vec{E}_s^{(f,b)}+\vec{E}_a^{(f,b)}$. For the symmetric mode we have
\begin{equation}
\vec{E}_s^{(f)} = \vec{E}_s^{(b)},
\end{equation}
but the antisymmetric mode exhibits odd parity and the scattered field in the forward direction is the negative of that in the backward direction,
\begin{equation}
\vec{E}_a^{(f)} = -\vec{E}_a^{(b)}.
\end{equation}
Hence, the symmetric field scatters equally in both directions whereas the antisymmetric mode scatters with opposite phase in the forward and backward directions. This can also be understood by considering the lattice as a uniform array of two-atom unit cells, with each unit cell of the symmetric mode consisting of an electric dipole moment with even forward/backward radiation parity.
The antisymmetric collective excitation has odd radiation parity. In a long-wavelength multipole description, its leading contributions are magnetic-dipole and electric-quadrupole moments with the same scattering parity.
Note, we use the labels $A$, $B$, to refer to the individual planes at $x=0$, $x=a$, respectively, while the labels $s$, $a$, refer to the modes consisting of symmetric and anti-symmetric polarization on each plane, respectively. This interference is illustrated schematically in Fig.~\ref{fig:sym}.

In the infinite-lattice limit, both modes have uniform polarization in the $yz$-plane, and hence the scattered light can be calculated by reducing the sum in Eqs.~(\ref{eq:Ef},\ref{eq:Eb}) to the sum over each unit cell.
Since only two modes are significantly driven, the response of the array is then described by the amplitude of the symmetric mode on each atom $\mathcal{P}_s$, and the amplitude of the antisymmetric mode on each atom, $\mathcal{P}_a$, with atoms in layer $A$ having amplitude $\mathcal{P}_a$ and atoms in layer $B$ having amplitude $-\mathcal{P}_a$. This leads to a simplified two-mode model in the place of the full \EQREF{eq:eom}
\begin{equation}
\label{eq:tmm}
\partial_t\begin{pmatrix}
\Pc_{s}  \\ \Pc_{a}
\end{pmatrix}=i(\Lambda +\delta\mathcal{H})\begin{pmatrix}
 \Pc_{s} \\ \Pc_{a}
\end{pmatrix}+\begin{pmatrix}
 \zeta_{s} \\  \zeta_{a}
\end{pmatrix}.
\end{equation}
Physically, \EQREF{eq:tmm} is a phase-matched two-resonance approximation for a subwavelength lattice. The incident field couples primarily to the modes with uniform in-plane phase, while all other collective modes have negligible overlap at normal incidence. The dominant resonances are the symmetric and antisymmetric modes shown in Fig.~\ref{fig:fig1}, whose different radiation parity controls the interference between the forward and backward fields.
The driving terms are given by (see Appendix~\ref{sec:tmmapp})
\begin{align}
\zeta_s &= i\sqrt{N}\mathcal{R}_0e^{ika/2}\cos{(ka/2}), \label{eq:driving1}\\
\zeta_a &= \sqrt{N}\mathcal{R}_0e^{ika/2}\sin{(ka/2)},\label{eq:driving2}
\end{align} 
where $\mathcal{R}_0=\mathcal{D}\mathcal{E}_0/\hbar$. The diagonal matrix 
$\Lambda = \mathrm{diag}(\lambda_{s},\lambda_{a})$ is given in terms of the collective resonance eigenvalues $\lambda_n=\delta_n+i\upsilon_n$. The matrix $\delta\mathcal{H}$ does not need to be diagonal 
and nondiagonal elements can induce couplings between the collective eigenmodes~\cite{Facchinetti16,Facchinetti18}. Here, we assume that level shifts only vary between the two layers and not between atoms in each layer, such that $\Delta^{(j)}=\Delta^{(A)}$ for $j=1\ldots N/2$ and $\Delta^{(j)}=\Delta^{(B)}$ for $j=N/2+1\ldots N$. Then
\begin{equation}
\label{eq:dh}
\setlength\arraycolsep{1pt}
\delta\mathcal{H}=\begin{pmatrix}
\tilde{\Delta} & \delta_{as}  \\
 \delta_{as}  & \tilde{\Delta}
\end{pmatrix},
\end{equation}
where 
\begin{align}
\label{eq:td1}
\tilde{\Delta} &= (\Delta^{(A)}+\Delta^{(B)})/2 =\Delta +(\delta^{(A)}+\delta^{(B)})/2, \\
\label{eq:bd2}
\delta_{as} &= (\Delta^{(A)}-\Delta^{(B)})/2 =(\delta^{(A)}-\delta^{(B)})/2.
\end{align} 
In fact, although we formulate the theory using the general case where the resonances between the layers may be shifted, we find that the best results in our model are achieved \emph{in the absence of any external level shifts}, i.e., when $\Delta^{(A)}=\Delta^{(B)}$, such that $\delta_{as}=0$. The Huygens' response is obtained from the natural symmetric and antisymmetric collective modes of an unshifted bilayer, rather than from externally imposed level shifts or a deliberately engineered multi-mode unit cell. Our numerical Huygens' surface demonstrations are consequently obtained by setting $\delta_{as}=0$ in all our examples.

The line shifts and linewidths of these two modes for a finite lattice are shown in Fig.~\ref{fig:evals}. Although the eigenvalues are shown for the specific value $n=30$, the dependence on $n$ is minor for all but the smallest lattices, with, e.g., line shifts differing by $\sim 6\%$ between $n=5$ and $n=40$ for $a=0.25\lambda$ and $d=0.91\lambda$, with almost identical qualitative dependence on $d$.

We next solve the steady state of \EQREF{eq:tmm}, given by $\dot{\mathbf{b}}=0$. We define the total shift of the eigenvalue from the laser frequency, including both any average level shift and the collective line shifts, as $Z_{s,a}=\lambda_{s,a}+\tilde{\Delta}$. Then we find
\begin{align}
    \mathcal{P}_s &= i\left(Z_a\zeta_s-\delta_{as}\zeta_a \right)/Z, \\
    \mathcal{P}_a &= i\left(Z_s\zeta_a -\delta_{as}\zeta_s \right)/Z,
\end{align}
where $Z=Z_sZ_a-\delta_{as}^2$. For equal atomic resonances between the atoms in the two layers we have $\delta_{as}=0$ and the expressions simplify to $\mathcal{P}_{s,a}=i\zeta_{s,a}/Z_{s,a}$.

Using the results of these analytical expressions, the scattered fields in the forward and backward direction can be calculated by adding the scattered field from each mode with appropriate phase, and are given by (Appendix~\ref{sec:tmmapp})
\begin{align}\label{eq:fieldsf}
\frac{\vec{E}^{(f)}}{\mathcal{E}_0} &=  \frac{2i\upsilon_{\mathrm{1D}}}{\mathcal{R}_0\sqrt{N}}[\Pc_s \cos{(ka/2)}\nonumber\\
&\qquad +i\Pc_a \sin{(ka/2)}]e^{ik(x-a/2)}\unitvec{e}_y, \\
\label{eq:fieldsb}
\frac{\vec{E}^{(b)}}{\mathcal{E}_0} &= \frac{2i\upsilon_{\mathrm{1D}}}{\mathcal{R}_0\sqrt{N}}[\Pc_s \cos{(ka/2)}\nonumber\\
&\qquad-i\Pc_a \sin{(ka/2)}]e^{-ik(x-a/2)}\unitvec{e}_y, 
\end{align}
respectively, where $\upsilon_{\mathrm{1D}}=3\pi\gamma/(k^2 d^2)$ is the linewidth of an ideal uniform mode for a single infinite (in the $yz$ plane) layer~\cite{CAIT}. Note here that since both $\mathcal{R}_0$ and $\mathcal{P}_{s,a}$ are directly proportional to $\mathcal{E}_0$, this ratio is independent of $\mathcal{E}_0$, and similarly is independent of $N$. 

\subsection{Optimal bilayer Huygens' surface}
\label{sec:hs}

The optimal bilayer Huygens' surface~\cite{Pfeiffer13,Decker15,Yu15,Chong15,Shalaev15} is obtained when the two phase-matched collective resonances are driven with equal effective strength and radiate forward amplitudes that are equal and backward amplitudes that have equal magnitude but opposite phases.
Then, far from resonance, light does not interact with the lattice and is transmitted with zero (equivalently $2\pi$) phase shift. On resonance, the symmetric and antisymmetric mode are equally excited, leading to destructive interference in the reflected field and constructive interference in the transmitted field with a $\pi$ phase shift, as in~\cite{Ballantine22bilayer}.

To achieve this equal occupation, the modes should be driven with equal strength, as occurs when the incident light has a $\pm\pi/2$ phase shift between the first and second layer, i.e.\ at $a=\lambda/4+j\lambda,3\lambda/4+j\lambda$, with $j$ integer [see Eqs.~\eqref{eq:driving1} and~\eqref{eq:driving2}]. As shown in Fig.~\ref{fig:evals} in a $n=30$ bilayer array, for $a=0.25\lambda$ and $a=1.25\lambda$ there are intralayer spacings $d\simeq 0.91\lambda$ and $d\simeq 0.96\lambda$, respectively, where the resonances of the two modes overlap. Hence, a Huygens' surface can be realized most simply at these spacings, where the in-plane uniform symmetric and anti-symmetric modes are both equally excited by the incident light (see Sec.~\ref{sec:tmm}). For $a=0.75\lambda$, the line shifts of the modes are always distinct, meaning there is no in-plane spacing $d$ where they can be driven equally, and we find no efficient solutions which give full $2\pi$ phase control without complex level shifts. 

While in our numerics we concentrate on the case of $a=0.25\lambda$, the two-mode model predicts identical results at $a=1.25\lambda$, as all fields are simply shifted by a full $2\pi$ phase shift, and this spacing may be easier to achieve experimentally. We vary $a$ and $d$ in Sec.~\ref{sec:spacing}, and find also other spacings where transmission is still reasonably high. 

\begin{figure}[t]
 \centering
  \includegraphics[width=\columnwidth]{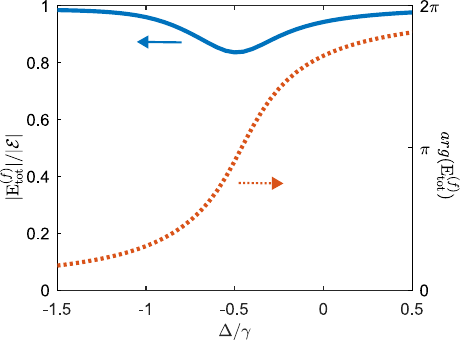}
 \caption{Amplitude (solid blue line, left axis, normalized to incident field) and phase (dashed red line, right axis, relative to incident field) of transmitted field through a bilayer Huygens' surface for an incident plane wave illumination as laser detuning is varied. Parameters are $a=0.25\lambda$, $d=0.91\lambda$, $n=30$; where the lattice consists of $n\times n$ atoms in the $yz$ plane and total atom number $2n^2$. }
 \label{fig:trans}
\end{figure}

Solving Eq.~(\ref{eq:tmm}) for the two cases with $a=0.25\lambda$ or $a=1.25\lambda$, when $\delta_s=\delta_a$ and $\upsilon_s=\upsilon_a$ (in the large lattice limit, these occur at $d/\lambda\simeq0.91$ and 0.97, respectively), we find in the absence of externally imposed level shifts on atoms
\begin{align}
\label{eq:pssolve}
\mathcal{P}_s &= -\sqrt{N}\mathcal{R}_0e^{i\pi/4}/(\sqrt{2}Z_s), \\
\label{eq:pasolve}
\mathcal{P}_a &= -\sqrt{N}\mathcal{R}_0e^{-i\pi/4}/(\sqrt{2}Z_{a}),
\end{align}
where $Z_s=Z_a$ for this spacing.
When $\unitvec{e}_\mathrm{in}=\unitvec{e}_d$, this trivially gives 
$\vec{E}^{(b)} = 0$, while for the numeric values of $\upsilon_{s,a}$ in an infinite array we find at the resonance $\tilde{\Delta}=-\delta_s$
\beq
\left|\frac{\vec{E}_{\rm tot}^{(f)}}{\mathcal{E}_0}\right|=\left|1-2\frac{\upsilon_{\mathrm{1D}}}{\upsilon_{s,a}}\right|\simeq1.
\label{eq:ratio}
\eeq
As the detuning is swept through the resonance from $\tilde{\Delta}\ll -\delta_s$ to $\tilde{\Delta}\gg -\delta_s$, the unwrapped transmission phase ${\rm arg}(E_{\rm tot}^{(f)})$  winds continuously by $2\pi$, while the transmission amplitude remains close to unity for all $\tilde\Delta$, i.e.\ representing an ideal Huygens surface. 
A numerical example for the incident plane-wave illumination is provided in Fig.~\ref{fig:trans}, showing the continuous $2\pi$ phase winding for $n=30$ as the laser detuning is varied,  with $|\vec{E}^{(f)}_\mathrm{tot}|/|\cbE|\agt 0.83$ for all detunings. The minimum transmission increases towards one for larger lattices, e.g.\ $|\vec{E}^{(f)}_\mathrm{tot}|/|\cbE| \gtrsim0.9$ for $n=50$. The forward scattering model can overestimate the decrease in transmission in small lattices, since coherently scattered light experiences a spread $\Delta k \sim 1/\sqrt{\cal A}$ in wavevector values that is not fully accounted for. Higher transmission values can be achieved in finite-size lattices by optimizing a Gaussian beam overlap with the density profile of the phase-coherent eigenmode~\cite{Facchinetti16,Facchinetti18}. We find for a Gaussian illumination $|\vec{E}^{(f)}_\mathrm{tot}|/|\cbE|  \agt 0.95$, with $n=30$, and $|\vec{E}^{(f)}_\mathrm{tot}|/|\cbE|  \agt 0.98$, with $n=50$.

The transmission minimum derived from Eqs.~(\ref{eq:pssolve},\ref{eq:pasolve}) and Eqs.~(\ref{eq:fieldsf},\ref{eq:fieldsb}) occurs exactly at the resonance for $\delta_s=\delta_a$ and $\upsilon_s=\upsilon_a$ and depends on the ratio $\upsilon_\mathrm{1D}/\upsilon_{a,s}$ according to \EQREF{eq:ratio}, approaching perfect transmission for $n\rightarrow \infty$. However, even for smaller lattices with $n\gtrsim 5$ there is an exact degeneracy where both the line shift and linewidth of the modes are equal and \EQREF{eq:ratio} remains valid within the forward scattering approximation. The leading finite size contribution is then only to change the transmission minimum through the ratio $\upsilon_\mathrm{1D}/\upsilon_{a,s}$ from the ideal one, and the qualitative behavior of the Huygens' surface is similar to the numerics presented here. 

The Huygens' surface can be used to produce arbitrary wavefront manipulation, e.g.\ orbital angular momentum, beam focussing, etc.\ \cite{Ballantine20Huygens,Ballantine21wavefront,Ballantine21quantum,Ballantine22str} by having the detuning $\Delta^{(j)}$ vary in the $yz$ plane, while remaining independent of $x$. However, it is worth noting that due to the relatively large in-plane spacing in the specific example case there is a limitation on how fast this can vary, which will limit, e.g.\ the maximal possible angle for beam-steering. 

\begin{figure}[t]
 \centering
  \includegraphics[width=\columnwidth]{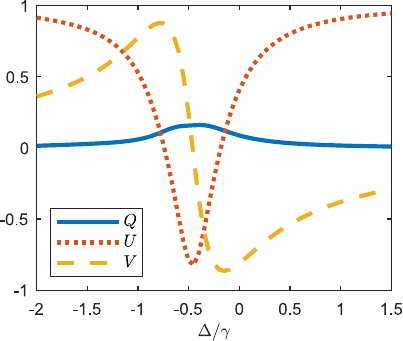}
 \caption{Polarization control via a bilayer Huygens' surface. The stokes parameters of the transmitted light, $Q$ (horizontal polarization), $U$ (diagonal polarization), and $V$ (circular polarization) are shown for an input plane wave with polarization $(\unitvec{e}_y+\unitvec{e}_z)/\sqrt{2}$, as a function of a spatially constant detuning $\Delta$ across the whole bilayer. As the phase of the $y$ component varies from 0 to $2\pi$ (as in Fig.~\ref{fig:trans}) relative to the unchanged $z$ component, the total polarization also varies. Parameters used are $n=30$, $d=0.91\lambda$, $a=0.25\lambda$. }
 \label{fig:pol}
\end{figure}

We numerically demonstrate an application of the bilayer Huygens' surface as a variable polarization wave-plate. The polarization of the transmitted light is encapsulated in the normalized Stokes parameters: $Q=(|E_{\mathrm{tot},z}|^2-|E_{\mathrm{tot},y}|^2)/|\mathbf{E}_{\mathrm{tot}}|^2$ corresponds to linear polarization in the $y$ or $z$ directions, $U=2\mathrm{re}(E_{\mathrm{tot,y}}E_{\mathrm{tot},z}^\ast)/|\mathbf{E}_{\mathrm{tot}}|^2$ corresponds to diagonal or anti-diagonal linear polatization, and $V=-2\mathrm{im}(E_{\mathrm{tot,y}}E_{\mathrm{tot},z}^\ast)/|\mathbf{E}_{\mathrm{tot}}|^2$ corresponds to left or right-circular polarizations. We assume the input light polarization  $\unitvec{e}_{\mathrm{in}}=(\unitvec{e}_y+ \unitvec{e}_z)/\sqrt{2}$. Since the atomic dipoles are fixed in the $y$ direction, the phase of the $y$ polarization component of the transmitted light varies with detuning, while the $z$ component is transmitted with constant phase. 
This means the total ouptut polarization, $(e^{i\phi}\unitvec{e}_y+ \unitvec{e}_z)/\sqrt{2}$ for a phase shift $\phi$, varies from diagonal  ($U=1$) to left-circular ($V=1$), anti-diagonal ($U=-1$), right-circular ($V=-1$) and back to diagonal. A numerical example is shown in FIg.~\ref{fig:pol}. In previous similar proposals~\cite{Ballantine20Huygens,Ballantine21wavefront,Ballantine21quantum,Ballantine22str}, the difference in the response of the array to $y$ and $z$ polarized light was due to the geometry of a more complex unit-cell as well as non-trivial atom-dependent level shifts. In contrast to these proposals, here we use a simple geometry of two square lattice layers without any spatially nonuniform external level shifts applied to the atoms, with the control obtained by the illuminating laser frequency alone. 

\subsection{Varying atomic spacing}
\label{sec:spacing}

\begin{figure}[t]
 \centering
  \includegraphics[width=\columnwidth]{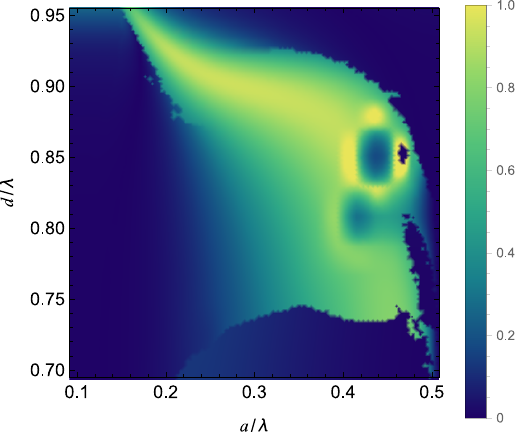}
 \caption{Absolute value $|E_{\rm tot}^{(f)}|/|\cbE|$ of forward transmitted field [derived from Eqs.~(\ref{eq:tmm},\ref{eq:fieldsf},\ref{eq:fieldsb}) using the collective linewidths and line shifts of the $n=30$ array] when on resonance with Huygens' surface for the 'worst-case' $\pi$ phase shift as function of $a$ and $d$, i.e.\ $|E_{\rm tot}^{(f)}|/|\cbE|=1$ corresponds to perfect transmission with maximum phase shift. Dark blue regions, even next to or contained within high-transmission regions, correspond to extremely low transmission.
 }
 \label{fig:eff1}
\end{figure}

Although the exact spacings of $a/\lambda=0.25$ and 1.25 allow for perfect interference between modes, small variations in the parameters will still lead to approximate solutions, where the constructive or destructive interferences are almost complete. The absolute value of the transmitted field when on resonance with the Huygens' surface inducing a $\pi$ phase shift, as predicted by the two-mode model, is shown in Fig.~\ref{fig:eff1} as a function of $a$ and $d$. While the two-mode model, Eq.~(\ref{eq:tmm}), is solved analytically, it requires the eigenvalues of each mode which are calculated numerically for a range of parameters, then interpolated for other values. Results are then derived by numerically optimizing the real part of the transmitted fields over $\tilde{\Delta}$ and $\delta_{as}$. Note this corresponds to the 'worst-case' scenario, where transmission is lowest, corresponding to a $\pi$ phase shift and the dip in transmission at the resonance in Fig.~\ref{fig:trans}. 

In the numerics, we allow for the relative level shifts of the atoms between the two layers to vary, while assuming them to remain equal in each individual layer. However, 
numerical optimization demonstrates that the transmission is maximized when $\delta_{as}=0$, i.e., when there are no externally imposed level shifts on atoms. Analytically, it is possible to show that $\delta_{as}=0$ is always a local minimum (or maximum) of the real part of the transmitted field. While there are also additional minima which may be real-valued, the global minimum when also varying $\tilde{\Delta}$ is always at $\delta_{as}=0$.

While the two-mode model predicts efficient solutions, this model is only accurate when these two modes dominate the response. 
The prediction of a two-mode model  in a $n=30$ array for $a=0.25\lambda$, $d=0.91\lambda$ with  $|\vec{E}^{(f)}_\mathrm{tot}|/|\cbE|\agt 0.95$ is reached for $n=30$ array simulations with a Gaussian beam illumination.
The condition for the ideal Huygens' surface solution demonstrated here is the overlapping mode resonances that can be simultaneously excited (although simultaneous excitation even without overlapping resonances can also be possible by using more complex level shifts in an isotropic $J=0\rightarrow J^\prime=1$ transition or more complex geometries).

\begin{figure}[t]
 \centering
  \includegraphics[width=\columnwidth]{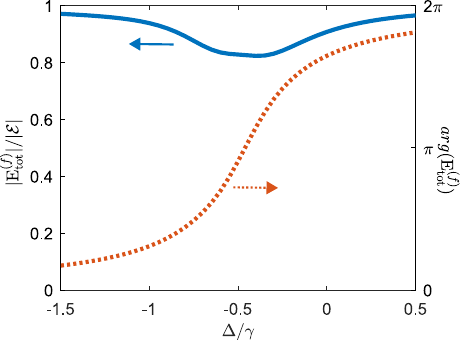}
 \caption{Amplitude (solid blue line, left axis, normalized to incident field) and phase (dashed red line, right axis, relative to incident field) of transmitted field for an incident plane wave illumination, for $a=0.3\lambda$, $d=0.90\lambda$, and $n=30$. While the minimum transmission $|\vec{E}^{(f)}_\mathrm{tot}|/|\cbE|\agt 0.82$ is only slightly reduced compared to Fig.~\ref{fig:trans}, this dip covers a broader range of phases.  }
 \label{fig:a3d9}
\end{figure}

Small changes in lattice spacings do not result in a significant reduction in efficiency, as can be verified from Fig.~\ref{fig:eff1} close to the optimal solution $a=0.25\lambda$, $d=0.91\lambda$ where the two-mode model remains accurate.
To illustrate that small variations in the lattice constants still provide a continuous full $2\pi$ phase control with high transmission, we also show  a specific numerical solution in Fig.~\ref{fig:a3d9} for $a=0.3\lambda$, $d=0.90\lambda$ using the same incident plane wave case as in  Fig.~\ref{fig:trans}. For a Gaussian incident beam, the transmission in this case increases to $\agt 0.92$. 
For a $n=50$ array $\agt 0.9$ can be reached even with an incident plane wave.

\subsection{Position fluctuations}

Estimating the effects of atomic position fluctuations is numerically more demanding. Mean-field approximations of fluctuations can be phenomenologically included by considering a reduced dipole radiation kernel due to the smearing of the atomic positions~\cite{castin09,Castin13,Perczel2017a}. Although this approach can qualitatively describe ensemble-averaged coherent transmission in multilayer arrays by shifting the collective line shift and linewidth towards their single atom values~\cite{ruks25}, it preserves the vanishing resonance splitting between the two relevant eigenmodes in a bilayer array and is insufficient to describe the position fluctuations for the Huygens’ surface. Consequently, we have analyzed representative cases using stochastic electrodynamics simulations~\cite{Ruostekoski2023,Lee16} by sampling atomic positions at each lattice site from a Gaussian density distribution $\sim\exp(-r^2/\eta^2)$  and averaging transmission over many realizations according to \EQREF{eq:Ef}. Because positional disorder scatters light into a broader range of transverse momenta, the forward-scattering model, which retains only the selected transmitted mode, is likely to overestimate the resulting degradation of transmission.
Nevertheless,  in the studied systems, we typically find 15--25\% reduction in the transmission minimum for $\eta=0.044 d$.

\section{Conclusions}

We have shown how atomic-array wavefront engineering can be reduced to a simple bilayer construction of two square arrays, which are not limited to small lattice spacings. The essential simplification is that the Huygens' surface response is generated by the intrinsic symmetric and antisymmetric collective modes of atoms in a simple lattice geometry without externally engineered differential level shifts between atoms in a unit cell or complex in-plane unit-cell structures. In advanced Huygens' surface applications, spatially varying wavefronts may then be produced by applying a common local detuning to both layers, without altering the internal Huygens' balance of the unit cell.

This minimal atomic Huygens' surface follows from collective scattering symmetry rather than from separately engineered electric- and magnetic-dipole resonances. The symmetric mode provides an electric-dipole-like 
scattering channel, while the antisymmetric mode supports the opposite forward/backward radiation parity through a simplified magnetic-dipole/electric-quadrupole-like excitation. Their interference cancels the reflected field 
and enables near-unity transmission while the unwrapped transmission phase winds continuously by $2\pi$ and is controlled only by the laser detuning. 
For atoms with an isotropic  $J=0\rightarrow J'=1$ transition, the same geometry can also be combined with the previously demonstrated subradiant storage mechanism (see Appendix~\ref{sec:iso}), in which ac-Stark shifts couple the radiating in-plane modes to long-lived out-of-plane modes. Thus the bilayer architecture provides a common platform for reflectionless wavefront shaping and, using the storage protocol of Ref.~\cite{Ballantine22bilayer}, single-photon storage and retrieval.

\subsection*{Acknowledgements}
We acknowledge financial support from the UK EPSRC, Grants No.\ EP/S002952/1 and No.\ EP/W005638/1. Data underlying the results presented in this paper are
available in~\cite{data}.

\appendix

\section{Derivation of two-mode model}
\label{sec:tmmapp}

To derive the two-mode model, we assume for simplicity an infinite lattice, i.e.\ $N\rightarrow\infty$, such that the eigenmodes of $\mathcal{H}$ are translationally invariant Bloch vectors with a uniform density. In this limit, $\mathcal{H}$ is a normal matrix: its eigenvectors become orthogonal and can be chosen real. More general cases are considered, e.g., in Ref.~\cite{Ruostekoski17}. We transform to a basis where $\mathcal{H}$ is diagonal, via an orthogonal matrix $O=\left[\mathbf{v}_1 \mathbf{v}_2 \ldots\right]^T$, whose row vectors, $\mathbf{v}_i$, are the eigenvectors of $\mathcal{H}$, including the symmetric and anti-symmetric modes. Applying this transformation to Eq.~~(\ref{eq:eom}), $\mathcal{H}^\prime=O\mathcal{H}O^T$ is a diagonal matrix of the eigenvalues of $\mathcal{H}$. However, $\delta\mathcal{H}^\prime=O\delta\mathcal{H}O^T$, as well as having diagonal terms which shift the resonances, will also have off-diagonal terms which couple different modes.  Furthermore, we define the amplitudes in this basis, $\mathbf{b}^\prime$ and the driving term; $\zeta=O\mathbf{f}$, 

In detail, we take the first two modes to be the ideal uniform (in the $yz$ plane) symmetric and anti-symmetric modes, respectively,
\begin{align}
    \mathbf{v}_s=\mathbf{v}_1 &=\left(1,1,\ldots,1,1,\ldots\right)/\sqrt{N}, \\
    \mathbf{v}_a=\mathbf{v}_2 &=\left(1,1,\ldots,-1,-1,\ldots\right)\sqrt{N},
\end{align}
up to a uniform global phase factor, where the first $N/2$ entries describe atoms at $x=0$, and the remaining terms describe atoms at $x=a$. 

For plane-wave illumination, the incident field is given by $\mathcal{E}(\mathbf{r})=\mathcal{E}_0e^{ikx}$. Then 
\begin{align}
    \vec{f} &= i\mathcal{R}_0 \left(1,1 \ldots e^{ika},e^{ika},\ldots\right) \\
        &= i\sqrt{N}\mathcal{R}_0\left[\mathbf{v}_1(1+e^{ika})+\mathbf{v}_2(1-e^{ika})\right]/2.
\end{align}

Since the rows of $O$ are $\mathbf{v}_j^T$, and using the orthogonality condition $\mathbf{v}_i^T\mathbf{v}_j=\delta_{ij},$ we see that 
\begin{align}
    \zeta_s &= \mathbf{v}_s^T\mathbf{f} = i\sqrt{N}\mathcal{R}_0e^{ika/2}\cos{(ka/2)}, \\
    \zeta_a &= \mathbf{v}_a^T\mathbf{f} = \sqrt{N}\mathcal{R}_0e^{ika/2}\sin{(ka/2)},
\end{align}
and $\zeta_j =0$, for all $j\neq 1,2$, due to the incident field only phase-matching two of the collective eigenmodes.

We derive the general formalism for the case of allowing for the relative level shifts of the atoms between the two layers (in the $x$ direction) to vary, while assuming them to remain equal in each individual layer (in the $yz$ plane).
The matrix $\delta\mathcal{H}=\mathrm{diag}(\mathbf{\Delta})$ reads
\begin{align}
    \mathbf{\Delta} &= \left(\Delta^{(A)},\ldots,\Delta^{(B)},\ldots\right).
\end{align}
Again using the biorthogonality of the eigenvectors, we 
\begin{multline}
    \delta\mathcal{H}^\prime_{ij}=O_{ik}\delta\mathcal{H}_{kl} O_{lj}^T \\ =\begin{cases}
        (\Delta^{(A)}+\Delta^{(B)})/2, & i=j=1,2 \\
        (\Delta^{(A)}-\Delta^{(B)})/2, & i,j=1,2;\: i\neq j \\
        0 & \text{otherwise}.
    \end{cases}
\end{multline}

The transformed amplitudes $\mathbf{b}^\prime_j=(O\mathbf{b})_j=\mathbf{v}_j^T\mathbf{b}$ simply give the polarisation amplitudes in the basis of eigenmodes, rather than a single atom basis. Since only two modes are driven, and these modes are only coupled to each other, the system of $N$ linear equations in Eq.~(\ref{eq:eom}) can be decomposed into the equations for these two modes, which govern the response of the lattice, and the remaining $N-2$ equations, which describe modes with zero amplitude in the steady state, leading to Eq.~(\ref{eq:tmm}). 

The scattered fields Eq.~(\ref{eq:fieldsf},\ref{eq:fieldsb}) are derived from Eq.~(\ref{eq:Ef}) by assuming that the atoms are in a superposition of the symmetric and antisymmetric modes, i.e., $\mathcal{P}^{(j)}=\mathcal{P}_s\vec{v}_1^{(j)}+\mathcal{P}_a\vec{v}_2^{(j)}$. Then, e.g., 
\begin{align}
\nonumber
    \sum_j &\mathcal{P}^{(j)}e^{ikx_j} = \frac{1}{\sqrt{N}}\big[ \sum_{j=1}^{N/2} (\mathcal{P}_s+\mathcal{P}_a)+\!\!\!\!\!\!\! \sum_{j=N/2+1}^{N} \!\!\!\!\!\! (\mathcal{P}_s-\mathcal{P}_a)e^{ika}\big] \nonumber \\
    &= \frac{\sqrt{N}}{2} \left[\mathcal{P}_s(1+e^{ika}) + \mathcal{P}_a(1-e^{ika})\right] \nonumber\\
    &=\sqrt{N} e^{ika/2}\left[\mathcal{P}_s \cos{(ka/2)}-i\mathcal{P}_a\sin{(ka/2)}\right],
\end{align}
and similarly for the sum over $\mathcal{P}^{(j)}e^{-ikx_j}$.

For a finite lattice, the vectors $\mathbf{v}_1$ $\mathbf{v}_2$ are not actually eigenmodes of $\mathcal{H}$ due to edge effects, and approach the uniform modes only as $N\rightarrow\infty$. However, due to phase matching, in reasonably-sized lattices the eigenmodes $\mathbf{v}_{s,a}$, with uniform phase in the $yz$ plane are still the dominant modes which couple to the incident field. While there is some small excitation of other modes, close to the resonance of $\mathbf{v}_{a,s}$ these are negligible, and we find the two-mode model to be an excellent approximation, especially for a Gaussian beam with waist comparable to the size of the lattice that matches the spatial profile of the modes~\cite{Facchinetti16,Facchinetti18}.

\section{Isotropic transition}
\label{sec:iso}

The results presented here concern two-level atoms with fixed dipole orientation. However,  similar results can be obtained with an isotropic $J=0\rightarrow J'=1$ transition, as found in alkaline-earth and rare-earth metals. In this case the polarization on each atom is described by three components, $\Pc^{(j)}_\nu$, where the dipole is given by
$\vec{d}_j=\sum_\nu\mathcal{D}\mathcal{P}^{(j)}_\nu\unitvec{e}_\nu$,
and $\nu=0,\pm 1$ correspond to the respective excited states, $\ket{J'=1;\nu=0,\pm 1}$. The dynamics can again be cast in matrix form, analogous to Eq.~\ref{eq:eom}, with the substitution $b_{3j+\nu-1}=\Pc^{(j)}_\nu$, 
$f_{3j+\nu+1}=i\mathcal{R}_\nu(\rv_j)$, $ {\cal R}_\nu=\mathcal{D}\pol_\nu^*\cdot\boldsymbol{\mathcal{E}}/\hbar$, and the off-diagonal elements of $\mathcal{H}$ given by~\cite{Ruostekoski2023}
\begin{equation}
\mathcal{H}_{3j+\nu-1,3k+\mu-1}=i\xi \unitvec{e}_\nu^{\ast}\cdot \mathsf{G}(\vec{r}_j-\vec{r}_k)\unitvec{e}_\mu.
\end{equation}

From numerical calculations, we find that normally-incident $y$-polarized input light will still predominantly excite modes which are polarized in the $y$ direction and approximately symmetric or anti-symmetric between the two layers, even though there is now some coupling to the $x$ and $z$ polarizations due to the finite size effects. For example, for a lattice of size $n=50$ and a $y$-polarized plane wave input, we find occupation of $L_{a,s}\gtrsim 0.4$ for each of the modes closest to an ideal uniform symmetric and uniform antisymmetric excitation, respectively, suggesting that over $80\%$ of the polarization is in the modes described by the two-mode model.

Introducing a relative level shift between the $m=\pm 1$ states, as could be achieved by magnetic fields or by ac Stark shifts of lasers or microwaves~\cite{gerbier_pra_2006}, induces a coupling between the excitations in the $y$ and $x$ directions even for an infinite lattice~\cite{Facchinetti16,Facchinetti18,Ballantine20ant,Parmee22b}. In Refs.~\cite{Ballantine21quantum,Ballantine22bilayer}, this technique was employed in storing single-photon pulses and in Ref.~\cite{Ballantine21PT} to generate exceptional points and PT symmetry breaking. The coupling is possible because dipoles pointing along the $y$ or $x$ direction no longer form an eigenbasis or evolve independently. Then the two-mode model described in Eq.~(\ref{eq:tmm}) can be generalised to a four-mode model, where the uniform in-plane symmetric and anti-symmetric modes are complimented by out-of-plane modes polarized in the $x$ direction, which again are approximately uniform in the $yz$ plane but either symmetric or anti-symmetric between the two layers. 

For completeness, we again consider a fairly general case of allowing for the relative level shifts of the atoms between the two layers (in the $x$ direction) to vary, while assuming them to remain equal in each individual layer (in the $yz$ plane).
The detunings  $\Delta_{j\nu}=\omega-\omega_{j\nu}$, from the level $\nu$ transition frequency $\omega_{j\nu}$,
 are then equal across each layer, with $\Delta^{(j)}_\mu=\Delta^{(A)}_\mu$ for $j\leq N/2$, with $N$ total atoms ordered such that $j\leq N/2$ refers to the bottom ($x=0$) layer, and $\Delta^{(j)}_\mu=\Delta^{(B)}_\mu$ for $j> N/2$, corresponding to the top ($x=a$) layer.  
Combining the transformation to Cartesian coordinates with that to a diagonal basis of $\mathcal{H}$, the full control can be described by a four-mode model:
\begin{equation}
\label{eq:fourmm2}
\partial_t\begin{pmatrix}
\Pc_{xa} \\ \Pc_{ya} \\ \Pc_{xs} \\ \Pc_{ys}
\end{pmatrix}=i(\Lambda + \delta\mathcal{H})\begin{pmatrix}
\Pc_{xa} \\ \Pc_{ya} \\ \Pc_{xs} \\ \Pc_{ys}
\end{pmatrix}+\begin{pmatrix}
0 \\ \zeta_{ya} \\ 0 \\ \zeta_{ys}
\end{pmatrix},
\end{equation}
with $\Lambda = \mathrm{diag}(\lambda_{xa},\lambda_{ya},\lambda_{xs},\lambda_{ys})$ for $\lambda_n=\delta_n+i\upsilon_n$ and
\begin{equation*}
\delta\mathcal{H}=\begin{pmatrix}
\tilde{\Delta} & -i\delta_{xy} & \delta_{as} & -i\delta_{as;xy} \\
i\delta_{xy} & \tilde{\Delta} & i\delta_{as;xy} & \delta_{as} \\
\delta_{as} & -i\delta_{as;xy} & \tilde{\Delta} & -i\delta_{xy} \\
i\delta_{as;xy} & \delta_{as} & i\delta_{xy} & \tilde{\Delta}
\end{pmatrix},
\end{equation*}
with 
\begin{align}
\label{eq:td2}
\tilde{\Delta} &= (\Delta_{+}^{(A)}+\Delta_{-}^{(A)}+\Delta_{+}^{(B)}+\Delta_{-}^{(B)})/4, \\
\label{eq:bd1}
\delta_{xy} &= (\Delta_{+}^{(A)}-\Delta_{-}^{(A)}+\Delta_{+}^{(B)}-\Delta_{-}^{(B)})/4, \\
\label{eq:bd4}
\delta_{as} &= (\Delta_{+}^{(A)}+\Delta_{-}^{(A)}-\Delta_{+}^{(B)}-\Delta_{-}^{(B)})/4, \\
\label{eq:td2}
\delta_{as;xy} &= (\Delta_{+}^{(A)}-\Delta_{-}^{(A)}-\Delta_{+}^{(B)}+\Delta_{-}^{(B)})/4,
\end{align} 
where the level shifts are uniform in the $yz$ plane but $A$ and $B$ superscripts refer to differing level shifts between the layers at $x=0$ and $x=a$.
Hence, $\tilde{\Delta}$ describes the resonance condition, $\delta_{xy}$ couples in-plane modes to out-of-plane modes while preserving symmetry~\cite{Facchinetti16,Facchinetti18,Ballantine22bilayer}, $\delta_{as}$ couples symmetric modes to antisymmetric modes while preserving direction, and $\delta_{as;xy}$ couples the symmetric in-plane to antisymmetric out-of-plane and vice-versa. The system may therefore allow for a richer phenomenology than the Huygens' surface operation for a two-level transition, potentially including optimal conditions for coupled eigenmodes. 

The out-of-plane modes in infinite arrays have zero linewidths and correspond to bound-state-in-a-continuum types of modes. In relevant sized finite arrays, they remain deeply subradiant, with decay rates orders of magnitude smaller than those of the in-plane modes~\cite{Facchinetti16,Facchinetti18,Ballantine20ant}.  They are therefore ideal for storing light, including single photons, as demonstrated for bilayer atomic arrays in Ref.~\cite{Ballantine22bilayer}. However, the incident light couples almost exclusively to the in-plane modes. 
Imposed  ac-Stark shifts that mix in-plane and out-of-plane polarizations can therefore be used to transfer the wavefront-shaped optical excitation into long-lived subradiant modes and later retrieve it. In this way, the present level-shift-free Huygens' mechanism and the established subradiant storage protocol can be integrated in the same bilayer architecture.

\end{document}